\documentclass[10pt,superscriptaddress,twocolumn,amsmath,amssymb,aps,pra]{revtex4-1}
\usepackage{mathrsfs}
\usepackage{graphicx}
\usepackage{dcolumn}
\usepackage{bm}

\renewcommand{\Re}{\operatorname{Re}}
\renewcommand{\Im}{\operatorname{Im}}

\newcommand{\be}{\begin{equation}}
\newcommand{\ee}{\end{equation}}
\newcommand{\bea}{\begin{eqnarray}}
\newcommand{\eea}{\end{eqnarray}}

\begin{document}
\title{Quantum Chaos of Unitary Fermi Gases in Strong Pairing Fluctuation Region}
\author{Xinloong Han}
\affiliation{Department of Physics and Center of Theoretical and Computational Physics, The University of Hong Kong, Hong Kong, China}
\author{Boyang Liu} \email{boyangleo@gmail.com} \affiliation{Institute of Theoretical Physics, Beijing University of Technology
Beijing 100124, China}

\date{\today}
\begin{abstract}
The growth rate of the out-of-time-ordered correlator in a N-flavor Fermi gas is investigated and the Lyapunove exponent $\lambda_L$ is calculated to the order of $1/N$. We find that the Lyapunove exponent monotonically increases as the the interaction strength increases from the BCS limit to the unitary region. At the unitarity the Lyapunove exponent increases while the temperature drops and it can reach to the order of $\lambda_L\sim T$ around the critical temperature for the $N=1$ case. The system scrambles faster for stronger pairing fluctuations. At the BCS limit, the Lyapunov exponent behaviors as $\lambda_L\propto e^{\mu/T} a^2_s T^2/N$.

\end{abstract}
 \maketitle

\section{Introduction}

Information scrambling is a crucial stage in thermalization of a closed system. During this process the quantum entanglement spreads across all the freedoms of the system and the memory of the initial state is lost, which is taken as a key prerequisite for thermalization. Recently, the studies in gauge gravity duality have inspired some new insights into the quantum chaos\cite{Susskind2008,Maldacena1999,Gubser,Witten,Shenker2014,Roberts,Shenker2015,Kitaev2014}. It is suggested the black holes are the fastest scramblers in nature\cite{Susskind2008}. Moreover, the experimental realizations of nearly isolated quantum systems  also attract increasing attention to this area \cite{Rigol,Langen,Kaufman}. Analogous to the Lyapunov exponents describing the growth of chaos in classical models, the scrambling to the quantum chaos can also be probed by growth rate of so called out-of-time-ordered correlator (OTOC).

\begin{figure}[h]
\includegraphics[width=0.3\textwidth]{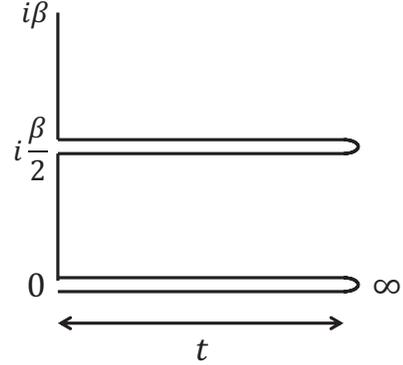}
\caption{The complex time contour for calculating the out-of-time-oredered correlators. The horizontal direction represents the real time evolution and the vertical direction respresents the imaginary time evolution. It contains two real time folds, which are seperated by $i\beta/2$.}
\label{fig:timecontour}
\end{figure}

The OTOC was first introduced by A. I. Larkin and Yu. N. Ovchinnikov in the study of superconductivity\cite{Larkin1969}. Recently, this subject is revived by the discovery of an unexpected bound on the Lyapunov exponent that is extracted from OTOC \cite{Susskind2008,Maldacena2016}. Several experiments on measurement of OTOC have been conducted \cite{zhu2016,Yao2016,Garttner2017,Li2017}. Usually, in stead of directly calculating the OTOC it's more convenient to evaluate the "regulated" squared commutator defined $\mathcal C(t)={\rm Tr}\{\sqrt \rho [W(t), V(0)]^\dagger \sqrt \rho[W(t), V(0)]\}$ \cite{Stanford,Chowdhury2017}, where $\rho=e^{-\beta H}$ is the thermal density matrix and $W$ and $V$ are  local Hermitian operators in general. It can be expanded as $\mathcal C(t)=2{\rm Tr}\{\sqrt\rho W(t)V(0)\sqrt\rho V(0)W(t)\}-2{\rm Re}[{\rm Tr}\{\sqrt\rho W(t)V(0)\sqrt\rho W(t)V(0)\}]$. The first term is time ordered. On the other hand, the second term is on an unusual time order as illustrated in Fig. \ref{fig:timecontour} and it's called OTOC. In a chaotic  system $\mathcal C(t)$ is expected to have an exponential behavior at the time scale $t_L$ as $\mathcal C(t)\sim  e^{\lambda_L t}$. Analogous to the classical chaos, $\lambda_L$ is called Lyapunov exponent and $t_L^{-1} \sim \lambda_L$. Based on some reasonable physical assumption, Lyapunov exponent is proven to have an upper bound of $2\pi k_BT/\hbar$ and it saturates in models with gravity duals\cite{Susskind2008,Maldacena2016}. An concrete example is the celebrated Sachdev-Ye-Kitaev (SYK) model\cite{Sachdev1993,Kitaev,Maldecena} which holds a conformal symmetry in the low-energy limit and is dual to an AdS$_2$ gravity theory.

In condensed matter physics, the systems usually don't possess conformal symmetry. However, there exist some exceptions. At the critical point the conformal symmetry can emerge for low energy and long distance. Investigations have been done in this regime \cite{Chowdhury2017,sachdevpnas2017,Yao2018}. In these system there are no quasi-particle excitations and the temperature is the only relevant scale. The Lyapunov exponents are found to obey the relationship of $\lambda_L\sim \kappa T$. The unitary Fermi gas is another  example with scaling invariance. With the properties of highly controllable and hyper clean it can be a perfect playground to investigate the information scambling\cite{Bentsen2019,Duan2019} and thermalization in closed quantum systems. At the unitary point, the non-relativistic conformal symmetry emerges and investigations have been taken to discuss its duality to a gravity theory\cite{McGreevy2008,D.T.Son2008}.  The behaviors of the Lyapunov exponent have been studied at both high temperature and low temperature limits \cite{P.Zhang2019}. However, it is more interesting to investigate the behavior around the critical temperature, where it has been shown more close to a non-Fermi liquid behavior\cite{Krinner2016,Liu2017,Husmann2018,Han2019}.

In this work, we calculate the Lyapunove exponent of a N-flavor Fermi gas with tunable interaction. The OTOC is evluated by a series of ladder diagrams and the Lyapunov exponent is calculated to the order of $1/N$.  As the interaction strength increases from the BCS limit to the unitary regime we find that the Lyapunov exponent monotonically increases while the temperature is fixed. We also investigate the temperature dependence of the Lyapunov exponent at the unitarity. $\lambda_L$ can increase to  $\lambda_L \sim T$ for $N=1$ case when the temperature is close to the critical temperature. Furthermore, we also find that the Lyapunov exponent behaves as $\lambda_L\propto z a_s^2T^2/N$ for high temperature at the BSC limit, where $a_s\rightarrow 0^-$.

\section{Model}
We will start from a system with N fermion flavors. The Hamiltonian can be cast as
\bea
\hat H=&&\int d^3{\bf r}\Big\{\sum_{i\sigma}\hat{\psi}^\dagger_{i\sigma}({\bf r})(-\frac{\nabla^2}{2m}-\mu)\hat\psi_{i\sigma}({\bf r})\cr &&-\frac{g}{N}\sum_{ij}\hat{\psi}^\dagger_{i\uparrow}({\bf r})\hat{\psi}^\dagger_{i\downarrow}({\bf r})\hat{\psi}_{j\downarrow}({\bf r})\hat{\psi}_{j\uparrow}({\bf r})\Big\},
\eea
where $\psi_{i\sigma}(\psi^{\dagger}_{i\sigma})$ is the  annihilation(creation) operator of the fermion field with flavor $i$ and spin $\sigma$. Parameter g is the bare interaction strength between the fermions. Here we assume the interaction strengths between different flavors are the same,  and it can be related to a s-wave scattering length $a_s$ by the following renormalization relation
\bea
\frac{1}{g}=-\frac{m}{4\pi a_s}+\int\frac{d^3\textbf{k}}{(2\pi)^3}\frac{1}{2\epsilon_{\textbf k}},\label{eq:gas}
\eea
where $\epsilon_{\textbf{k}}=k^2/2m$, and $m$ is the mass of the fermions. By introducing an auxiliary bosonic field $\varphi$ the four-fermion interaction term can be decoupled through the Hubbard-stratonovich transformation. Then in the imaginary time path integral formulism the partition function  can be written as $\mathcal{Z}=\int\mathcal{D}[\psi_{i\sigma},\psi^{\dagger}_{i\sigma},\varphi,\bar{\varphi}]e^{-S[\psi_{i\sigma},\psi^{\dagger}_{i\sigma},\varphi,\bar{\varphi}]}$, where the action $S$ is
\bea
&&S[\psi_{i\sigma},\psi^{\dagger}_{i\sigma},\varphi,\bar{\varphi}]=\nonumber \\
&&\int d \tau d^3 {\bf r}\Big(\sum_{i\sigma}\psi^{\dagger}_{i\sigma}(\tau,{\bf r})(\partial_\tau-\frac{\nabla^2}{2m}-\mu)\psi_{i\sigma}(\tau,{\bf r})-\nonumber \\
&&\sum_{i}\varphi\psi^{\dagger}_{i\uparrow}\psi^{\dagger}_{i\downarrow}-\sum_{i}\bar{\varphi}\psi_{i\downarrow}\psi_{i\uparrow}+\frac{ N\bar{\varphi}\varphi}{g} \Big).
\eea In this  work we set $\hbar=1$.
	
The imaginary time Greens' functions of fermion and boson are defined as $\delta_{ij}\delta_{\sigma\sigma'}G(\tau,\textbf{r})=\langle\psi^\dagger_{i\sigma}(\tau,\textbf{r}) \psi_{j\sigma'}(0,{0})\rangle$ and $\mathcal{G}(\tau,\textbf{r})=\langle \bar\varphi(\tau,\textbf{r}) \varphi(0,0)\rangle$, respectively. In the momentum space the free propagators can be simply expressed as
\bea
&&G^{(0)}(i\omega^f_n,{\bf k})=\frac{1}{i\omega_n-\epsilon_{\bf k}+\mu},\cr &&
\mathcal G^{(0)}(i\omega^b_n,{\bf k})=g/N,\label{eq:ImGF}
\eea
where $\omega^f_n=(2n+1)\pi/\beta$ and $\omega^b_n=2n\pi /\beta$ are the Matsubara frequencies for fermions and bosons, respectively, and $\beta=1/k_B T$. In order to calculate the Lyapunov exponent up to the order of $1/N$ we will involve the dressed propagators of fields $\psi$ and $\varphi$ as shown in Fig. \ref{fig:feynman}. The dressed propagator of $\varphi$ is a resummation of bubble diagram. Then, it's written as
\bea
\mathcal G(i\omega^b_n,{\bf k})=\frac{g/N}{1-g\Pi(i\omega^b_n,{\bf k})},\label{eq:fullphi}
\eea
where $\Pi(i\omega^b_n,{\bf k})$ is the one-loop bubble
\bea
&&\Pi(i\omega^b_n,\textbf{q})=\int\frac {d^3 \textbf{k}}{(2\pi)^3}\frac{1-n_F(\epsilon_{\bf k}-\mu)-n_F(\epsilon_{\bf q-\textbf k}-\mu)}{-i\omega^b_n+\epsilon_{\bf k}+\epsilon_{\textbf q-\textbf k}-2\mu}.\cr&&
\eea
$n_F(\epsilon_{\bf k}-\mu)=1/\exp(\beta(\epsilon_{\bf k}-\mu)+1)$ is the Fermi-Dirac distribution function.
The dressed propagator of field $\psi_i$ is
\bea G(i\omega^f_n,\textbf k)=\frac{1}{-i\omega^f_n+\epsilon_{\bf k}-\mu-\Sigma(i\omega^f_n,\textbf{k})}, \label{eq:fullpsi}
\eea
where the self-energy of fermions $\Sigma(i\omega^f_n,\vec{k})$ is expressed as
\bea
&&\Sigma(i\omega^f_n,\textbf{k})=\frac{1}{\beta}\sum_{\omega^b_m}\int\frac{d^3 \textbf{q}}{(2\pi)^3} \frac{\mathcal{G}(i\omega^b_m,\textbf{q})}{-i\omega^b_m+i\omega^f_n+\epsilon_{\textbf {q}-\textbf {k}}-\mu}.\cr&& \eea

The corresponding retarded Green's functions are defined as usual as $\delta_{ij}\delta_{\sigma\sigma'}G_R(t,\textbf{r})=-i\theta(t)\langle \{\psi_{i\sigma}(t,\textbf{r}),\psi_{j\sigma'}^{\dagger}(0,0)\}\rangle$ and $\mathcal{G}_R(t,\vec{r})=-i\theta(t)\langle [\varphi(t,\vec{r}),\bar{\varphi}(0,0)]\rangle$, where $\theta(t)$ is the heaviside step function. In momentum space the forms of the retarded Green's functions can be obtained by the analytic continuation of the Eq.(\ref{eq:fullphi}) and (\ref{eq:fullpsi}) as
$\mathcal G_R(\omega,{\bf k})=\mathcal G(i\omega^b_n\rightarrow\omega+i0^+,{\bf k})$ and $G_R(\omega,{\bf k})=G(i\omega^f_n\rightarrow\omega+i0^+,{\bf k})$. Then $G_R(\omega,{\bf k})$ is written as
\bea
G_R(\omega,\textbf k)=\frac{1}{-\omega-i 0^+ +\epsilon_{\bf k}-\mu-\Sigma(\omega+i0^+,\textbf{k})}.
\eea Hence, in the dressed retarded Green's function the pole is modified by the self-energy. Working to the first order in $\Sigma$ the pole can be approximately calculated as $\omega^\ast=\epsilon_{\bf k}-\mu-\Re \Sigma(\epsilon_{\bf k}-\mu+i0^+,\textbf{k})+i\Gamma(k)$, and the quantum scattering rate $\Gamma(k)$ is defined as $\Gamma(k)\equiv-\Im \Sigma(\epsilon_{\bf k}-\mu+i0^+,\textbf{k})$.

In order to evaluate the OTOC we need to define the symmetrized Wightman function as
\bea
&&\delta_{ij}\delta_{\sigma\sigma'}G_W(t,{\bf r})={\rm Tr}\{\sqrt\rho\psi_{i\sigma}(t,{\bf r})\sqrt\rho\psi^\dagger_{i\sigma}(0,0)\},\cr&&
\mathcal G_W(t,{\bf r})={\rm Tr}\{\sqrt\rho\varphi(t,{\bf r})\sqrt\rho\bar\varphi(0,0)\}.
\eea
In the momentum space they can be written in terms of the spectral functions of fields $\psi_{i\sigma}$ and $\varphi$ as
\bea
&&G_W(\omega, {\bf k})=\frac{A_F(\omega, {\bf k})}{2\cosh(\omega\beta/2)},\cr&&
\mathcal G_W(\omega, {\bf k})=\frac{A_B(\omega, {\bf k})}{2\sinh(\omega\beta/2)},
\eea
The spectral functions can be calculated as the imaginary parts of the retarded Green's functions, $A_F(\omega,\textbf{k})=-2 \Im {G}_R(\omega,\textbf{k})$ and $A_B(\omega,\textbf{k})=-2 \Im \mathcal{G}_R(\omega,\textbf{k})$.

\section{The Lyapunov exponent}
\begin{figure}[t]
\includegraphics[width=0.48\textwidth]{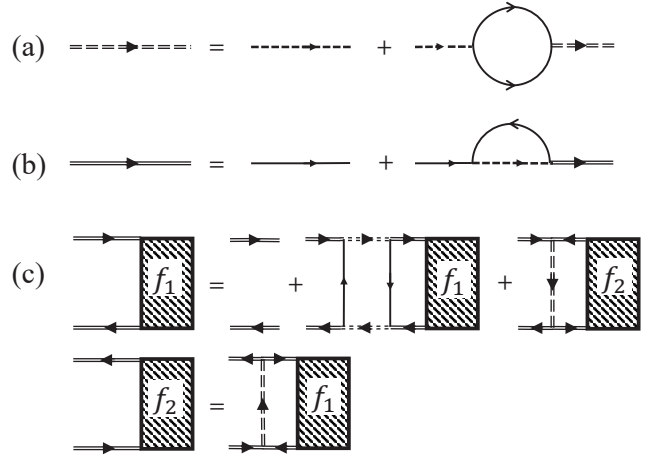}
\caption{(a) The Feynman diagram of the Dyson-Schwinger equation for field $\varphi$. (b) The Feynman diagram of the Dyson-Schwinger equation for field $\psi_{i\sigma}$. The double solid (dashed) line and the solid (dashed) line represent the dressed and free propagators of $\psi_{i\sigma}(\varphi)$, respectively. (c) The Feynman diagrams of the Bethe-Salpeter equations of the squared anti-commutators. }
\label{fig:feynman}
\end{figure}

In order to calculate the Lyapunov exponent it's convenient to evaluate the "regulated" squared anti-commutator defined as \cite{Chowdhury2017,Yao2018}
\bea
\mathcal C_{1}(t)=&&\frac{\theta(t)}{N^2}\sum_{i,j} \int d^3\textbf{r}{\rm Tr}\Big[\sqrt{\rho}\{\psi_{i\uparrow}(t,\textbf{r}),\psi_{j\uparrow}^{\dagger}(0,0)\} \cr &&
\times\sqrt{\rho}\{\psi_{i\uparrow}(t,\textbf{r}),\psi_{j\uparrow}^{\dagger}(0,0)\}^{\dagger}\Big].
\eea
The factor $1/N^2$ is to normalized the summation of indices $i,j$. Since the system is symmetric about exchanging spin indice, without losing any generality we investigate the "regulated" squared anti-commutator of field $\psi_{i\uparrow}$ as above. For the calculation up to the order of $1/N$ the squared anti-commutator $\mathcal C_1$ will couple to another squared anti-commutator $\mathcal C_2$ as demonstrated in Fig. \ref{fig:feynman} (c). The squared anti-commutator $\mathcal C_2$ is written as the following
\bea
\mathcal C_{2}(t)=&&\frac{\theta(t)}{N^2}\sum_{i,j} \int d^3\textbf{r}{\rm Tr}\Big[\sqrt{\rho}\{\psi_{i\downarrow}^{\dagger}(t,\textbf{r}),\psi_{j\uparrow}^{\dagger}(0,0)\}\cr &&\times\sqrt{\rho}\{\psi_{i\downarrow}^{\dagger}(t,\textbf{r}),\psi_{j\uparrow}^{\dagger}(0,0)\}^{\dagger}\Big]. \eea
At the moment of $t=0$ the above anti-commutators vanish because of ${\bf r}\neq 0$. However, in chaotic system the time evolution of the operators may involve increasing degree of freedoms. As a result the fields become nonlocal at later time. It is conjectured that the squared anti-commutators will have an exponential growth $\mathcal C_i(t)\sim e^{\lambda_L t}$ at short time. Analogously to the approach in ref. \cite{Stanford}, in order to compute the $\lambda_L$ to the leading order in $1/N$ we only keep the fastest-growing diagrams, which is a set of ladder diagrams as shown in Fig.\ref{fig:feynman} (c). The "rails" of the ladder correspond to the retarded Green's functions. They are defined on the two real time folds. The two rails are separated by an imaginary time difference $i\beta/2$ and they are connected by "rungs". The "rungs" correspond to the Wightman Green's functions.

The Fourier transformation of $\mathcal C_i(t)$ is denoted as $\mathcal C_i(\omega)$ with $\mathcal C_i(t)=\int d \omega e^{-i\omega t} \mathcal C_i(\omega)$. To sum up all the ladder series it's convenient to define functions $f_i(\nu;\omega,{\bf k})$ as
\bea
\mathcal C_i(\nu)=\frac{1}{N}\int\frac{d\omega d^3{\bf k}}{(2\pi)^4}f_i(\nu;\omega,{\bf k}).
\eea
The lowest order of $f_1(\nu;\omega,{\bf k})$ is simply expressed as $G_{R}(\omega, {\bf k})G^\ast_R(\omega-\nu,\textbf{k})$. Summation of all the ladder diagrams yields the Bethe-Salpeter equations
\bea
f_1(\nu;\omega,{\bf k})=&&G_{R}(\omega,{\bf k})G^{*}_R(\omega-\nu,\textbf{k})\Big(1+\int\frac{d\omega^{\prime}d^3 {\bf k}^{\prime}}{(2\pi)^4}  \cr &&
\big(\mathcal{K}_1(\nu;\omega,{\bf k};\omega^{\prime},{\bf k}^{\prime})f_2(\nu;\omega^{\prime},{\bf k}^{\prime})\cr&&+\mathcal{K}_2(\nu;\omega,{\bf k};\omega^{\prime},{\bf k}^{\prime})f_1(\nu;\omega^{\prime},{\bf k}^{\prime})\
\big) \Big),
\cr f_2(\nu;\omega,{\bf k})=&&G_{R}(\omega,{\bf k})G^{*}_R(\omega-\nu,\textbf{k})\cr &&\int\frac{d\omega^{\prime}d^3 {\bf k}^{\prime}}{(2\pi)^4} \mathcal{K}_1(\nu;\omega,{\bf k};\omega^{\prime},{\bf k}^{\prime})f_1(\nu;\omega^{\prime},{\bf k}^{\prime}),\cr&&\label{eq:BSequation}
\eea
where $\mathcal{K}_1$ and $\mathcal{K}_2$ are the integral kernels corresponding to the one-rung and two-rung diagrams in Fig.\ref{fig:feynman} (c), respectively. They are written as
\bea
&&\mathcal{K}_1(\nu;\omega,{\bf k};\omega^{\prime},{\bf k}^{\prime})=\mathcal{G}_W(\omega^\prime+\omega,{\bf k}+{\bf k}^{\prime}),\cr &&
\mathcal{K}_2(\nu;\omega,{\bf k};\omega^{\prime},{\bf k}^{\prime})=\cr&&~~~~~~N\int\frac{d\omega^{\prime\prime}d^3\textbf{k}^{\prime\prime}}{(2\pi)^4}\mathcal{G}_R(\omega^{\prime\prime},{\bf k}^{\prime\prime})\mathcal{G}_R^*(\omega^{\prime\prime}-\nu,\textbf{k}^{\prime\prime})\cr&&~~~~~~
\times G_W(\omega+\omega^{\prime\prime},{\bf k}+{\bf k}^{\prime\prime}) G_W(\omega^\prime+\omega^{\prime\prime},{\bf k}^{\prime}+{\bf k}^{\prime\prime}).\cr&&
\eea

For the following calculation we will take several approximations. Firstly, one expects that the $f_1(\nu;\omega,{\bf k})$ to be exponentially growing, while first term of $f_1$ in Eq. (\ref{eq:BSequation}) will be decaying. Hence, this term can be safely dropped without affecting the evaluation of the growth rate. Secondly, the pair of fermionic Green's functions $G_{R}(\omega,\textbf{k})G_R^\ast(\omega-\nu,\textbf{k})$ in Eq. (\ref{eq:BSequation}) can be approximated as $\frac{2\pi i\delta(\omega-\epsilon_{\textbf{k}}+\mu)}{\nu+2i\Gamma(k)}$. Thirdly, because in the above approximation all pairs of the retarded Green's functions include a on-shell delta function, it's natural to postulate the on-shell form of $f_i(\nu;\omega, {\bf k})$ as $f_i(\nu;\omega,{\bf k})\approx f_i(\nu;\textbf{k})\delta(\omega-\epsilon_{\textbf{k}}+\mu)$\cite{Stanford,Chowdhury2017}. Please refer to the appendix A for the details of the approximation.  With all above approximations the Bethe-Saltpeter equations of Eq.(\ref{eq:BSequation}) can be reduced to
\bea
&&(-i\omega+2T\tilde{\Gamma}(\tilde k))f_1(\omega;{\tilde k})=\cr&&~~~~~~~\frac{T}{N}\int \frac{d \tilde k^{\prime} \tilde k^{\prime}}{\tilde k}\Big(\tilde{\mathcal{K}}_{1}(\tilde k,\tilde k^{\prime})f_2(\omega;{\tilde k}^{\prime})+\tilde{\mathcal{K}}_2(\tilde k,\tilde k^{\prime})  f_1(\omega;{\tilde k}^{\prime}) \Big),\cr
&&(-i\omega+2T\tilde{\Gamma}(\tilde k))f_2(\omega;{\tilde k})=\frac{T}{N}\int \frac{d \tilde k^{\prime} \tilde k^{\prime}}{\tilde k}\tilde{\mathcal{K}}_{1}(\tilde k,\tilde k^{\prime})f_1(\omega;{\tilde k}^{\prime}),\cr &&\label{eq:BSequation2}
\eea
where the momenta have been rescaled to be dimensionless as $\tilde k= k/\sqrt{T}$ and $\tilde k^{\prime}=k^{\prime}/\sqrt{T}$. Correspondingly we define a dimensionless quantum scattering rate $\tilde \Gamma =\Gamma/T$. Here we have assumed the function $f(\omega,{\bf k})$ is rotationally invariant and integrated over the angles. Then the function $f_i(\omega,{\bf k})$ is reduced to $f_i(\omega, k)$ in Eq. (\ref{eq:BSequation2}) . The dimensionless functions $\tilde{\mathcal{K}}_1$ and $\tilde{\mathcal{K}}_2$ are written as
\bea
&&\tilde{\mathcal{K}}_1(\tilde k,\tilde k^{\prime})=N\int_{|\tilde k^{\prime}-\tilde k^{\prime\prime}|}^{\tilde k^{\prime}+\tilde k^{\prime\prime}} \frac{\tilde p d\tilde p}{(2\pi)^2}\tilde{\mathcal{G}}_W(\tilde \epsilon_{k^{\prime}}+\tilde \epsilon_{k^{\prime\prime}}-2\tilde \mu,\tilde p),
\cr&&
\tilde{\mathcal{K}}_2(\tilde k,\tilde k^{\prime})=N^2\int \frac{d \tilde k^{\prime\prime}d \tilde \omega^{\prime\prime}}{128\pi^5} \frac{|\tilde{\mathcal{G}}_R(\tilde \omega^{\prime\prime},\tilde k^{\prime\prime})|^2 \Theta(\tilde k,\tilde k^{\prime},\tilde k^{\prime\prime})}{\cosh(\frac{\tilde  \epsilon_k-\tilde \mu-\tilde \omega^{\prime\prime}}{2})\cosh(\frac{\tilde \epsilon_{k^{\prime}}-\tilde \mu-\tilde \omega^{\prime\prime}}{2})},\cr&&
\label{eq:Kappa}\eea
where $\tilde k^{\prime\prime}=k^{\prime\prime}/\sqrt T$, $\tilde \omega^{\prime\prime}=\omega^{\prime\prime}/T$, $\tilde  \epsilon_k=\epsilon_k/T$ and $\tilde \mu=\mu/T$ and the bosonic retarded Green's functions and Wightman function are also rescaled to be dimensionless by $\tilde{\mathcal{G}}_R=\sqrt T\mathcal{G}_R$ and $\tilde{\mathcal{G}}_W=\sqrt T\mathcal{G}_W$. The $\Theta$ function is defined as $\Theta(k,k^{\prime},k^{\prime\prime})=\theta(2k^{\prime}k^{\prime\prime}+\epsilon_{\textbf{k}^{\prime\prime}}-\mu+k^{\prime\prime 2})\theta(2k^{\prime}k^{\prime\prime}-\epsilon_{\textbf{k}^{\prime\prime}}+\mu-k^{\prime\prime 2})\nonumber \theta(2k k^{\prime\prime}+\epsilon_{\textbf{k}^{\prime\prime}}-\mu+k^{\prime\prime 2})\theta(2k k^{\prime\prime}-\epsilon_{\textbf{k}^{\prime\prime}}+\mu-k^{\prime\prime 2})$.

To more easily solve for the Lyapunov exponent the Bethe-Saltpeter equations of Eq. (\ref{eq:BSequation2}) can be written in a simply form
\bea
&&-i\omega \mathcal F(\omega;{\tilde k})=\frac{T}{N}\int d \tilde k^{\prime} \mathcal{S}(\tilde k,\tilde k^{\prime})\mathcal F(\omega;\tilde k^{\prime}).\cr &&\label{eq:BSequation3}
\eea
where $\mathcal F^{T}(\omega;{\tilde k})=(\tilde k f_1(\omega;{\tilde k}),\tilde k f_2(\omega;{\tilde k}))$ and the dimensionless integral kernel $\mathcal{S}(\tilde k,\tilde k^{\prime})$ is defined as following
\bea
&&\mathcal{S}(\tilde k,\tilde k^{\prime})= \cr &&\left(\begin{matrix}
\tilde{\mathcal{K}}_2(\tilde k,\tilde k^{\prime})-2N\tilde{\Gamma}(\tilde k^{\prime})\delta(\tilde k-\tilde k^{\prime}) && \tilde{\mathcal{K}}_1(\tilde k,\tilde k^{\prime}) \cr
\tilde{\mathcal{K}}_1(\tilde k,\tilde k^{\prime}) && -2N\tilde{\Gamma}(\tilde k^{\prime})\delta(\tilde k-\tilde k^{\prime})
\end{matrix}\right). \cr && \label{eq:kernel}
\eea

We do not know how to solve Eq. (\ref{eq:BSequation3}) analytically. However, it can be solved numerically by discretizing the momenta $\tilde k$ and $\tilde k^\prime$ in the integral kernel $\mathcal{S}(\tilde k,\tilde k^{\prime})$. Then, the integral becomes the summation over the discrete momentum and Eq. (\ref{eq:BSequation3}) can be written as
\bea
-i\omega \mathcal F(\omega;{\tilde k}_i)=\frac{T}{N}\sum_{\tilde k_j}\mathcal{S}(\tilde k_i,\tilde k_j)\mathcal F(\omega;\tilde k_j),
\eea where $\tilde k_i$ is the discrete momentum with a small internals. Obviously $-i\omega$ is given by the eigenvalues of the kernel $\mathcal{S}(\tilde k_i,\tilde k_j)$ multiplied by a factor $T/N$. The Lyapunov exponent corresponds to the largest eigenvalue. Please refer to the appendix B for the details of the numerical calculation of the Lyapunov exponent.

\section{Quantum Chaos at the unitary point}
\begin{figure}[t].
\includegraphics[width=0.48\textwidth]{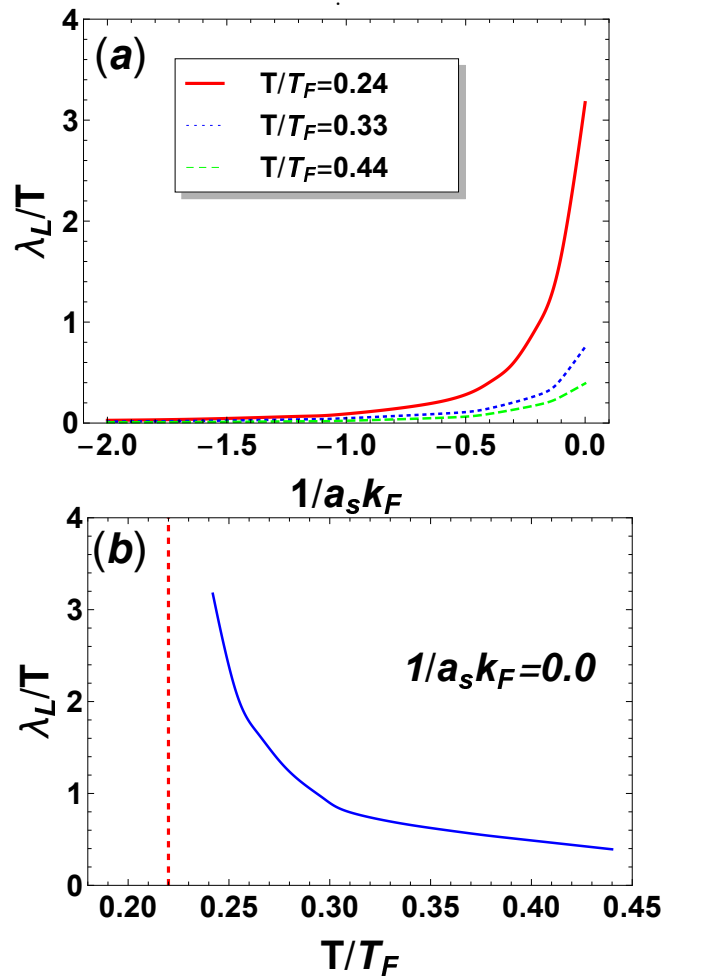}
\caption{(Color online) (a) $\lambda_L/T$ as a function of $1/a_sk_F$. The red solid, the blue dotted and the green dashed curves correspond to different temperatures $T/T_{F}=0.24,0.33$ and $0.44$, respectively. (b) $\lambda_L/T$ as a function of temperature $T/T_F$ for case of $1/a_sk_F=0$.}
\label{fig:unitary}
\end{figure}
In this section we study the case of unitary Fermi gases by setting $N=1$. This is not a fully controllable choice. However, since we only focus on the variations of the Lyapunov exponent with respect to the scattering length $a_s$ and the temperature, it may generate qualitative correct interpretation as the large N cases and inspire useful insight. In Fig. \ref{fig:unitary} (a) we plot $\lambda_L/T$ as a function of $1/a_sk_F$ for fixed temperature $T/T_{F}=0.24,0.33$ and $0.44$. If we compare these temperatures with the critical temperature calculated in the Nozi\`{e}res and Schmitt-Rink (NSR) scheme \cite{NSR,Ohashi}, which is $T_{c}=0.22T_F$ at $1/a_sk_F=0$, they can be written as $T/T_{c}=1.1, 1.5$, and $2.0$. One observes that the Lyapunov exponent monotonically increases as $1/a_sk_F$ goes from the BCS limit to the unitary regime. For lower temperature the $\lambda_L$ increases much faster the the higher temperature cases. At the unitary point $1/a_sk_F=0$ we polt $\lambda_L/T$ as a function of temperature $T/T_{F}$ in Fig. \ref{fig:unitary} (b). As the temperature drops the Lyapunov exponent monotonically increases and approaches the upper bound $2\pi T$. At the temperature of $T/T_{F}=0.24$, which corresponds to $T/T_c=1.1$ the Lyapunov exponent can reach a value of $\lambda_L\approx 3.2 T$. Here we would like to point out that we won't be able to expolre the region very close to $T_c$, where our numerical calculation becomes unstable since the propagator of Eq. (\ref{eq:fullphi}) diverges at $T_c$.

At the unitary point and temperature close to the critical point the system possesses two features. Firts, the system is scaling invariant. It obeys the non-relativistic conformal symmetry (the Schr\"{o}dinger group). Investigations have been taken for the possible non-relativistic version of ADS/CFT duality \cite{McGreevy2008,D.T.Son2008}. Second, it has been shown that around the critical temperature the system demonstrate a behavior of non-Fermi liquid due to the strong pairing fluctions \cite{Krinner2016,Liu2017,Husmann2018,Han2019}. several researches has shown that certain systems lacking of quasi-particle excitations demonstrate strong chaos \cite{Swingle2013,Chowdhury2017,sachdevpnas2017,Yao2018,Sachdev2018RMP}. Hence, it's not surprising that our system scrambles the fastest at the unitarity and the temperature close to Tc.

\section{The behaviors in the BCS limit}
At the BCS limit the scattering length $a_s\rightarrow0^-$. The retarded Green's function $\mathcal G_R$ of the field $\varphi$ can be expanded in terms of small $a_s$ as the following
\bea
&&\mathcal G_R(\omega, {\bf k})\cr&&=\frac{1/N}{-\frac{m}{4\pi a_s}+\int\frac {d^3 \textbf{k}}{(2\pi)^3}\frac{1}{2\epsilon_{\bf k}}-\int\frac {d^3 \textbf{k}}{(2\pi)^3}\frac{1-n_F(\epsilon_{\bf k}-\mu)-n_F(\epsilon_{\bf q-\textbf k}-\mu)}{-\omega-i0^++\epsilon_{\bf k}+\epsilon_{\textbf q-\textbf k}-2\mu}}\cr&&
\propto a_s/N.\eea
Notice that the temperature must be far from the critical temperature. Otherwise, according to the Thouless criterion one has $-\frac{m}{4\pi a_s}+\int\frac {d^3 \textbf{k}}{(2\pi)^3}\frac{1}{2\epsilon_{\bf k}}-\int\frac {d^3 \textbf{k}}{(2\pi)^3}\frac{1-n_F(\epsilon_{\bf k}-\mu)-n_F(\epsilon_{\bf q-\textbf k}-\mu)}{-\omega+\epsilon_{\bf k}+\epsilon_{\textbf q-\textbf k}-2\mu}\rightarrow 0$ when $T$ approaches $T_c$ and $\mathcal G_R$ can not be expanded for small $a_s$. Furthermore, since the Wightman function $\mathcal G_W$ the quantum scattering rate $\Gamma(k)$ can be calculated by $\mathcal{G}_W(\omega,\textbf{k})=\frac{-\Im \mathcal{G}_R(\omega,\textbf{k})}{\sinh(\omega\beta/2))}$ and $\Gamma({k})=\int\frac{d^3 {\bf q}}{2(2\pi)^3}\mathcal{G}_W(\epsilon_{\bf k}+\epsilon_{\bf {q}-{k}}-2\mu,{\bf q}) \frac{\cosh((\epsilon_{\bf k}-\mu)/{2T})}{\cosh((\epsilon_{\bf {q}-{k}}-\mu)/{2T})}$, their behaviors for small $a_s$ can be easily derived as $\mathcal G_R\propto z a_s^2\sqrt T/N$ and $\Gamma(k)\propto z a_s^2 T^2/N$, where $z\equiv \exp(\mu/T)$ is the fugacity. Please refer to Appendix C for the details. $\tilde{\mathcal{K}}_1(\tilde k, \tilde k^\prime)$ and $\tilde{\mathcal{K}}_2(\tilde k, \tilde k^\prime)$ in the integral kernel of Eq. (\ref{eq:kernel}) are functions of $\mathcal G_R$ and $\mathcal G_W$ as shown in Eq. (\ref{eq:Kappa}). Then it's straight forward to obtain the behaviors as $\tilde{\mathcal{K}}_1(\tilde k, \tilde k^\prime)\propto za^2_s T$ and $\tilde{\mathcal{K}}_2(\tilde k, \tilde k^\prime)\propto za^2_s T$. The three terms in Eq. (\ref{eq:kernel}) are all have the same asymptotic form of $za^2_s T$. Hence, as $a_s\rightarrow 0^-$ the Lyapunov exponent behaves as $\lambda_L\propto za_s^2 T^2/N$, which is consistent with the results on the Fermi liquid theory with well defined quasi-particles \cite{Igor2016,Banerjee,P.Zhang2019}.

\section{Conclusions}
We have computed the Lyapunov exponent for a N-flavor Fermion system using $1/N$ expansion. The variation of the Lyapunov exponent with respect to the scattering length $a_s$ and the temperature $T$ has been investigated. When $T$ is fixed the Lyapunov exponent monotonically increases as the $1/a_sk_F$ increases from the BCS limit to the unitary regime. When the scattering length is fixed to $1/a_sk_F=0$ the Lyapunov exponent increases while the temperature drops. Around the critical temperature it can reach to the order of $\lambda_L\sim T$ for $N=1$ case. Basically, our results indicate that with strong pairing fluctations the system exhibits strong chaos. Furthermore, the behavior of $\lambda_L$ at the BCS limit was calculated as $\lambda_L\propto z a_s^2 T^2/N$, which is consistent with the Fermi liquid theory.

\section{Acknowledgements}
We thank Shizhong Zhang, Yu Chen and Pengfei Zhang for very helpful discussions. The work is supported by the National Science Foundation of China (Grant No. NSFC-11874002), Beijing Natural Science Foundation (Grand No. Z180007) and Hong Kong Research Grants Council, GRF 17304719, CRF C6026-16W and C6005-17G.

\appendix
\section{Approximations for the reduction of Eq. (\ref{eq:BSequation2})}
With the first approximation the first term of $f_1$ in Eq. (\ref{eq:BSequation}) is dropped. Then the the Bethe-Salpeter equations in Eq. (\ref{eq:BSequation}) is reduced to
\bea
f_1(\nu;\omega,{\bf k})=&&G_{R}(\omega,{\bf k})G^{*}_R(\omega-\nu,\textbf{k})\int\frac{d\omega^{\prime}d^3 {\bf k}^{\prime}}{(2\pi)^4}  \cr &&
\big(\mathcal{K}_1(\nu;\omega,{\bf k};\omega^{\prime},{\bf k}^{\prime})f_2(\nu;\omega^{\prime},{\bf k}^{\prime})\cr&&+\mathcal{K}_2(\nu;\omega,{\bf k};\omega^{\prime},{\bf k}^{\prime})f_1(\nu;\omega^{\prime},{\bf k}^{\prime})\
\big),
\cr f_2(\nu;\omega,{\bf k})=&&G_{R}(\omega,{\bf k})G^{*}_R(\omega-\nu,\textbf{k})\cr &&\int\frac{d\omega^{\prime}d^3 {\bf k}^{\prime}}{(2\pi)^4} \mathcal{K}_1(\nu;\omega,{\bf k};\omega^{\prime},{\bf k}^{\prime})f_1(\nu;\omega^{\prime},{\bf k}^{\prime}),\cr&& \label{eq:BSappen}
\eea The second approximation is performed on the pair propagators $G_{R}(\omega,{\bf k})G^{*}_R(\omega-\nu,\textbf{k})$. In the free fermion case it's expressed as
\bea
&&G_{R}(\omega,{\bf k})G^{*}_R(\omega-\nu,\textbf{k})=\cr&&\frac{1}{\omega-\epsilon_{\bf k}+\mu+i0^+}\frac{1}{\omega-\nu-\epsilon_{\bf k}+\mu-i0^+}.
\eea
The integration over $\omega$ can be evaluated by the method of residue. Then it's straight forward to yield
\bea
&&G_{R}(\omega,{\bf k})G^{*}_R(\omega-\nu,\textbf{k})=\frac{2\pi i\delta(\omega-\epsilon_{\bf k}+\mu)}{\nu+2i0^+}.
\eea
The approximation is taken by replacing the $0^+$ by the scattering rate $\Gamma(k)$ for the interacting case. Then,
\bea
&&G_{R}(\omega,{\bf k})G^{*}_R(\omega-\nu,\textbf{k})=\frac{2\pi i\delta(\omega-\epsilon_{\bf k}+\mu)}{\nu+2i\Gamma(k)}.
\eea
As discussed in the maintext the third approximation is to postulate the on-shell form $f_i(\nu;\omega,{\bf k})\approx f_i(\nu;\textbf{k})\delta(\omega-\epsilon_{\textbf{k}}+\mu)$. Then the Eq. (\ref{eq:BSappen}) can be written as
\bea
&&(-i\nu+2\Gamma(k))f_1(\nu;{\bf k})= \int\frac{d\omega^{\prime}d^3 {\bf k}^{\prime}}{(2\pi)^4}2\pi\delta(\omega-\epsilon_{\textbf{k}}+\mu)\cr&&
\Big(\mathcal{K}_1(\nu;\omega,{\bf k};\omega^{\prime},{\bf k}^{\prime})f_2(\nu;{\bf k}^{\prime}) +\mathcal{K}_2(\nu;\omega,{\bf k};\omega^{\prime},{\bf k}^{\prime})f_1(\nu;{\bf k}^{\prime})\Big),\cr&&
(-i\nu+2\Gamma(k))f_2(\nu;{\bf k})= \int\frac{d\omega^{\prime}d^3 {\bf k}^{\prime}}{(2\pi)^4}2\pi\delta(\omega-\epsilon_{\textbf{k}}+\mu)\cr&&
\mathcal{K}_1(\nu;\omega,{\bf k};\omega^{\prime},{\bf k}^{\prime})f_1(\nu;{\bf k}^{\prime}).
\eea
Assuming $f_i(\nu;{\bf k}^{\prime})$ is rotationally invariant and performing the integration by implementing the delta function $\delta(\omega-\epsilon_{\textbf{k}}+\mu)$ one obtains the Eq. (\ref{eq:BSequation2}).

\section{Remarks on Numerical technique}
\begin{figure}[h]
\includegraphics[width=0.42\textwidth]{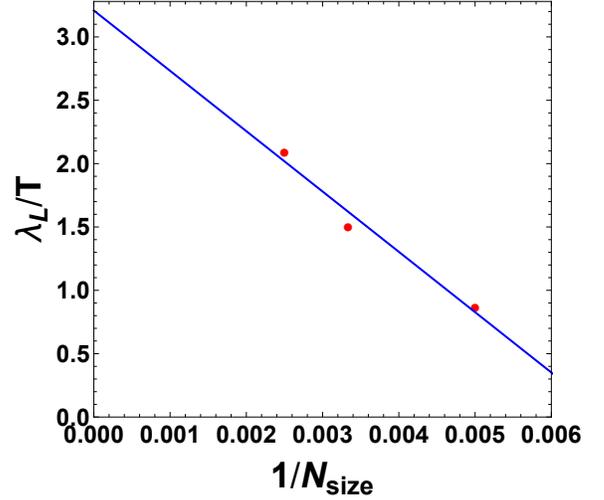}
\caption{(Color online) The extrapolation of $\lambda_L/T$ as a function of the discretized interval $1/N_{size}$. This plot is for the case of $1/a_sk_F=0$ and $T/T_F=0.24$.}
\label{fig:FindLyp}
\end{figure}

To numerically solve for the Lyapunov exponent we first discretize the momenta $\tilde k$ and $\tilde k^\prime$ of the integral kernel $\mathcal{S}(\tilde k,\tilde k^{\prime})$ in Eq. (\ref{eq:kernel}) into $N_{size}$ pieces. The cutoffs of momenta $\tilde k$ and $\tilde k^\prime$ are set to $\Lambda=15$. We have also checked the convergence of the results by performing the calculation for larger cutoffs. The kernel $\mathcal{S}(\tilde k,\tilde k^{\prime})$ is symmetric for exchanging $\tilde k$ and $\tilde k^\prime$. Then it can be easily diagonalized to obtain the eigenvalues, which are denoted as $\lambda_i$ here.  The Lyapunov exponent is related to the largest eigenvalue as $\lambda_L N/T={\rm max}(\lambda_i)$. Then the same calculation is performed for different $N_{size}$ and the corresponding value of $\lambda_L N/T$ is obtained. As an example we illustrate the case of $1/a_sk_F=0$ and $T/T_F=0.24$ in Fig. \ref{fig:FindLyp} . The final value of $\lambda_L N/T$ is read by the extrapolation to $1/N_{size}=0$.

\section{Behaviors at BCS limit}
At the BCS limit one has $a_s^{-1}\rightarrow-\infty$. Then the asymptotic behaviors of various propagators and the scattering rate $\Gamma(k)$ are demonstrated as the following. The full propagator of field $\varphi$ is
\bea
\mathcal{G}_R(\omega,\textbf{k})=\frac{1/N}{1/g-\Pi(\omega,\textbf{k})}\equiv\frac{1/N}{{Re}+i {Im}},
\eea
where
\bea
{Re}=&&-\frac{m}{4\pi a_s}+\int\frac {d^3 \textbf{k}}{(2\pi)^3}\frac{1}{2\epsilon_{\bf k}}\cr&&-\int\frac {d^3 \textbf{k}}{(2\pi)^3}\frac{1-n_F(\epsilon_{\bf k}-\mu)-n_F(\epsilon_{\bf q-\textbf k}-\mu)}{-\omega+\epsilon_{\bf k}+\epsilon_{\textbf q-\textbf k}-2\mu} \cr{Im}=&&-\pi\int\frac {d^3 \textbf{k}}{(2\pi)^3}(1-n_F(\epsilon_{\bf k}-\mu)-n_F(\epsilon_{\bf q-\textbf k}-\mu))\cr&&\delta(-\omega+\epsilon_{\bf k}+\epsilon_{\textbf q-\textbf k}-2\mu).
\eea
After we rescale all the momenta and frequency by ${\bf k}\rightarrow k/\sqrt{T}$, ${\bf q}\rightarrow {\bf q}/\sqrt{T}$ and $\omega\rightarrow \omega/T$ it's straight forward to get the following asymptotic behaviors for large $a_s^{-1}$
\bea
&&{Re}\propto a_s^{-1},\cr&&
Im \propto \sqrt T.
\eea
Notice that the temperature here must be far from the superfluid critical temperature, otherwise $Re\rightarrow 0$.
Then for large $a_s^{-1}$ the propagator $\mathcal{G}_R(\omega,\textbf{k})$ behaves as
\bea
\mathcal{G}_R(\omega,\textbf{k})\propto a_s/N.
\eea
The imginary part of $\mathcal{G}_R(\omega,\textbf{k})$ is
\bea
\Im \mathcal{G}_R(\omega,\textbf{k})=-\frac{1}{N}\frac{Im}{Re^2+Im^2}\propto a_s^2 \sqrt{T}/N.
\eea
The Wightman function of field $\varphi$ behaves as
\bea
\mathcal{G}_W(\omega_k-2\mu,\textbf{k})&&\equiv\frac{A_B(\omega_k-2\mu,\textbf{k})}{2\sinh((\omega_k-2\mu)\beta/2)}\cr&&=\frac{-\Im \mathcal{G}_R(\omega_k-2\mu,\textbf{k})}{\sinh((\omega_k-2\mu)\beta/2))}\cr&&\propto z a_s^2 \sqrt{T}/N. \label{eq:appGW}
\eea

The self-energy of fermions is
\bea
&&\Sigma(i\omega^f_n,\textbf{k})=\frac{1}{\beta}\sum_{\omega^b_m}\int\frac{d^3 \textbf{q}}{(2\pi)^3} \frac{\mathcal{G}(i\omega^b_m,\textbf{q})}{-i\omega^b_m+i\omega^f_n+\epsilon_{\textbf {q}-\textbf {k}}-\mu},\cr&& \eea
where the summation over $\omega^b_m$ is equivalent to a contour integration as the following
\bea
\Sigma(i\omega^f_n,\textbf{k})=&&\int\frac{d^3 \textbf{q}}{(2\pi)^3}\Big(\int\frac{dz}{2\pi i}\frac{n_B(z)(\mathcal{G}_R(z,{\bf q})-\mathcal{G}_A(z,{\bf q}))}{-z+i\omega^f_n+\epsilon_{\bf q -k}-\mu} \cr&&-\mathcal{G}(i\omega^f_n+\epsilon_{\bf {q}-{k}}-\mu,{\bf q})n_F(\epsilon_{\bf {q}-{k}}-\mu)\Big),
\eea
where $\mathcal{G}_A$ is the advanced Green's function for field $\varphi$. After we take a analytical continuation the imaginary part of the self-energy can be calculated as
\bea
&&\Im \Sigma(\omega+i0^+,{\bf k})\cr&&
=-\int\frac{d^3{\bf q}}{2(2\pi)^3}  \Bigg( n_F(\epsilon_{\bf {q}-{k}}-\mu){A}_B(\omega+\epsilon_{\bf {q}-{k}}-\mu)\cr&&~~~+\int d z {A}_B(z)
\delta(-z+\omega+\epsilon_{\bf {q}-{k}}-\mu) n_B(z)\Bigg) \cr&&=-\int\frac{d^3{\bf q}}{2(2\pi)^3}\mathcal{G}_W(\omega+\epsilon_{\bf {q}-{k}}-\mu,{\bf q})\frac{\cosh(\frac{\omega}{2T})}{\cosh(\frac{\epsilon_{\bf {q}-{k}}-\mu}{2T})}.\cr&&
\eea

The quantum scattering rate is defined as $\Gamma(k)=-\Im \Sigma(\epsilon_{\bf k}-\mu+i0^+,\textbf{k})$.
Then it can be written as
\bea
&&\Gamma({k})=\int\frac{d^3 {\bf q}}{2(2\pi)^3}\mathcal{G}_W(\epsilon_{\bf k}+\epsilon_{\bf {q}-{k}}-2\mu,{\bf q}) \frac{\cosh(\frac{\epsilon_{\bf k}-\mu}{2T})}{\cosh(\frac{\epsilon_{\bf {q}-{k}}-\mu}{2T})}.\cr&&
\eea
As we have derived in Eq. (\ref{eq:appGW}) the asymptotic behavior of the Wightman function is $\mathcal{G}_W(\epsilon_{\bf k}+\epsilon_{\bf {q}-{k}}-2\mu,{\bf q})\propto z a_s^2 \sqrt{T}/N$, then the asymptotic behavior of the quantum scattering rate for large $a_s^{-1}$ is as the following
\bea
\Gamma({k})\propto z a_s^2 T^2/N.
\eea
With all above asymptotic forms of $\mathcal{G}_R(\omega,\textbf{k})$, $\mathcal{G}_W(\omega,\textbf{k})$ and $\Gamma(k)$ straight forward calculation yields
\bea
&&\tilde{\mathcal{K}}_1(\tilde k, \tilde k^\prime)\propto z a_s^2T,\cr&&
\tilde{\mathcal{K}}_2(\tilde k, \tilde k^\prime)\propto z a_s^2T,
\eea and hence
\bea
\mathcal{S}(\tilde k,\tilde k^{\prime})\propto z a_s^2T.
\eea

Then the asymptotic behavior of Laypunov exponent $\lambda_L$ for large $a_s^{-1}$ is
\bea
\lambda_L \propto T (z a_s^2 T)/N=z a_s^2 T^2/N.
\eea


\begin{thebibliography}{Chaos}
\bibitem{Susskind2008} Y. Sekino, L. Susskind, J. High Energy Phys. \textbf{2008}, 065 (2008).
\bibitem{Maldacena1999} J. Maldacena, Advances in Theoretical and Mathematical Physics, vol. 2, no. 2, pp. 231, 1998, [International Journal of Theoretical Physics, vol. 38, article 1113, 1999].
\bibitem{Gubser} S. S. Gubser, I. R. Klebanov, and A. M. Polyakov, Physics Letters B,
vol. 428, no. 1-2, pp. 105-114, 1998.
\bibitem{Witten} E. Witten, Anti de Sitter space and holography, Advances in
Theoretical and Mathematical Physics, vol. 2, no. 2, pp. 253-291,
1998.
\bibitem{Shenker2014} S.H. Shenker, D. Stanford, J. High Energy Phys. \textbf{2014}, 067 (2014).
\bibitem{Roberts} D. A. Roberts, D. Stanford, and L. Susskind, Localized shocks,
Journal of High Energy Physics, vol. 1503, no. 51, 2015.
\bibitem{Shenker2015} S.H. Shenker, D. Stanford, J. High Energy Phys. \textbf{2015}, 132 (2015).
\bibitem{Kitaev2014} A. Kitaev, Hidden correlations in the hawking radiation
and thermal noise, talk given at Fundamental Physics Prize
Symposium, in Proceedings of the Stanford SITP seminars, 2014.
\bibitem{Rigol} M. Rigol, V. Dunjko, and M. Olshanii, Thermalization
and its mechanism for generic isolated quantum systems,
Nature (London) 452, 854 (2008).
\bibitem{Langen} T. Langen, R. Geiger, M. Kuhnert, B. Rauer, and J.
Schmiedmayer, Local emergence of thermal correlations
in an isolated quantum many-body system, Nat. Phys. 9,
640 (2013).
\bibitem{Kaufman} A. M. Kaufman, M. E. Tai, A. Lukin, M. Rispoli, R.
Schittko, P. M. Preiss, and M. Greiner, Quantum ther-
malization through entanglement in an isolated many-
body system, Science 353, 794 (2016).
\bibitem{Larkin1969} A. I. Larkin,  and Yu. N. Ovchinnikov, J. Exp. Theor. Phys. \textbf{28}, 1200 (1969).
\bibitem{Maldacena2016}J. Maldacena, S. H. Shenker, and D. Stanford, J. High Energy Phys. \textbf{08}  106 (2016).
\bibitem{zhu2016} G. Zhu, M. Hafezi, and T. Grover, Phys. Rev. A \textbf{94}, 062329 (2016).
\bibitem{Yao2016} N. Y. Yao, F. Grusdt, B. Swingle, M. D. Lukin, D. M. Stamper-Kurn, J. E. Moore, E. Demler, arXiv:1607.01801(2016).
\bibitem{Garttner2017} M. G\"{a}ttner, J. G. Bohnet, A. Safavi-Naini, M. L. Wall, J. J. Bollinger, and A. M. Rey, Nat. Phys. \textbf{13}, 781 (2017).
\bibitem{Li2017} J. Li, R. Fan, H. Wang, B. Ye, B. Zeng, H. Zhai, X. Peng, and J. Du, Phys. Rev. X. \textbf{7}, 031011 (2017).
\bibitem{Stanford} D. Stanford, J. High Energy Phys. \textbf{10}, 1007 (2016).
\bibitem{Chowdhury2017} D. Chowdhury, and B. Swingle, Phys. Rev. D. \textbf{96}, 065005 (2017).
\bibitem{Sachdev1993} S. Sachdev, and J. Ye, Phys. Rev. Lett. \textbf{70}, 3339 (1993).
\bibitem{Kitaev} A. Kitaev, A simple model of quantum holography. KITP http://online.kitp. ucsb.edu/online/entangled15/kitaev/ (2015).
\bibitem{Maldecena} J. Maldacena, D. Stanford, Phys. Rev. D. \textbf{94}, 106002 (2016).
\bibitem{sachdevpnas2017} Aavishkar A. Patel and Subir Sachdev, Proc. Natl. Acad. Sci. \textbf{114}, 1844 (2017).
\bibitem{Yao2018} S. Jian, and H. Yao, arXiv:1805.12299 (2018).
\bibitem{Bentsen2019}G. Bentsen, T. Hashizume, A. S. Buyskikh, E. J. Davis, A. J. Daley, S. S. Gubser, and M. Schleier-Smith, Phys. Rev. Lett. \text{123}, 130601 (2019).
\bibitem{Duan2019} C. B. Da\v{g}, and L.-M. Duan, Phys. Rev .A \textbf{99}, 052322 (2019).
\bibitem{McGreevy2008} K. Balasubramanian, and J. McGreevy, Phys. Rev. Lett. \textbf{101}, 061601 (2008).
\bibitem{D.T.Son2008} D. T. Son, Phys. Rev. D \textbf{78}, 046003 (2008).
\bibitem{P.Zhang2019} P. Zhang, Journal of Physics B: Atomic, Molecular and Optical Physics \textbf{52}, 13 (2019).
\bibitem{Krinner2016} S. Krinner, M. Lebrat, D. Husmann, C. Grenier, J.-P. Brantut, and T. Esslinger, Proc. Natl. Acad. Sci. USA {\bf 113}, 8144 (2016).
\bibitem{Liu2017} B. Liu, H. Zhai, and S. Zhang, Phys. Rev. A \textbf{95}, 013623 (2017).
\bibitem{Husmann2018} D. Husmann, M. Lebrat, S. H\"{a}usler, J.-P. Brantut, L. Corman, and T. Esslinger, Proc. Natl. Acad. Sci. {\bf 115}, 8563 (2018).
\bibitem{Han2019} X. Han, B. Liu, and J. Hu,  Phys. Rev. A {\bf 100}, 043604 (2019).
\bibitem{NSR} P. Nozi\`{e}res and S. Schmitt-Rink, J. Low Temp. Phys. \textbf{59}, 195 (1985).
\bibitem{Ohashi} Y. Ohashi, and A. Griffin, Phys. Rev. Lett. \textbf{89}, 130402 (2002).

\bibitem{Swingle2013} B. Swingle, and T. Senthil, Phys. Rev. B. \textbf{87}, 045123 (2013).
\bibitem{Sachdev2018RMP} S. A. Hartnoll, A. Lucas, and S. Sachdev, 	arXiv:1612.07324 (2018).

\bibitem{Igor2016} Igor L. Aleiner, L. Faoro, and Lev B. Ioffe, Annals of Physics \textbf{375}, 378 (2016)



\bibitem{Banerjee} S. Banerjee and E. Altman, Phys. Rev. B {\bf 95}, 134302 (2017).
\end{thebibliography}
\end{document}